\documentclass[prd,aps,showpacs,epsf,floats,onecolumn]{revtex4}
\usepackage[utf8x]{inputenc}
\usepackage{graphicx}
\usepackage{amsfonts}
\usepackage{amsmath}
\usepackage{amssymb}
\usepackage{hyperref}
\usepackage{bookmark}
\usepackage{xcolor}
\usepackage{soul}

\allowdisplaybreaks

\begin{document}

\title{\textbf{Upper Bounds on Fluctuation Growths of Observables in Open Quantum Systems}}

\author{\textbf{Newshaw Bahreyni}$^{1}$, \textbf{Paul M. Alsing}$^{2}$, \textbf{Carlo Cafaro}$^{2}$, \textbf{Walid Redjem}$^{2}$, \textbf{Christian Corda}$^{3,4}$} 
\affiliation{$^{1}$Pomona College, Claremont, CA 91711, USA}   
\affiliation{$^{2}$University at Albany-SUNY, Albany, NY 12222, USA}  
\affiliation{$^{3}$SUNY Polytechnic Institute, Utica, NY 13502, USA}
\affiliation{$^{4}$Department of Physics and
Electronics, Christ University, Bengaluru, 560029, Karnataka, India}

\begin{abstract}

The upper bounds for the rate of fluctuation growth of an observable in both open and closed quantum systems have been studied actively recently.  In our recent work we showed that the rate of fluctuation growth for an observable in a closed quantum system is upper bounded by the fluctuation of its corresponding velocity-like observable.  That bound also indicated a tradeoff between the time derivatives of the mean and the standard deviation.  In this paper we will look at open quantum systems in two cases.  For the first case we find the generator of evolution for an open system employing both the Taylor expansion and the standard time-ordered evolution via the Dyson series, while in the second case we consider no specific information about the evolution of the system.  We then find the rate of fluctuation growth in each case.  Comparing the upper bounds for each case and considering the upper bound found for a closed system suggest that including more details by separating the contributions of the system and state dynamics seems to result in looser bounds for the rate of fluctuation growth.  
   
\end{abstract}

.

\maketitle

\section{Introduction}
Time evolution of a closed quantum system is well known under the time-dependent Schr\"{o}dinger equation which explains that the evolution of a system's state under its Hamiltonian is governed by a unitary time-evolution operator given by $U(t)$. This is Schrodinger's well-known picture for temporal evolution \cite{Sch}.  This means that the probability is conserved and there is no loss of information which indicates that the entropy remains constant.  Equivalently, the system can be described in terms of the density matrix and its evolution.   In a closed quantum system in state $\vert \psi \rangle$ evolution of the average value $\langle A \rangle$ of an observable $A(t)$ that in general can be time-dependent is given by 
\begin{equation}
\label{eq:1}
\frac{d\langle A\rangle}{dt}=\left\langle  \partial_{t}A\right\rangle +i\langle \left[ \mathrm{H},A\right] \rangle.
\end{equation}
where $\hbar$ is set equal to 1 \cite{Eh, Sa, Coh}, and $\partial_{t}A=\partial A/\partial t$. 
For an observable $A$, its average value is given by 
\begin{equation}
\label{eq:1.1}
\langle A\rangle=tr(\rho A)=\Sigma_{j}p_{j}\langle \psi_{j}\vert A\vert \psi_{j}\rangle
\end{equation} 
where, in general, $\rho$ is the density operator that describes the state of knowledge of a quantum system .  Note that for a pure state $\rho^{2}=\rho$ whereas for a mixed state $\rho^{2}\neq \rho$.  Time evolution of $\rho$ depends on whether the system is open or closed.  
In general, when a system is closed and in a pure state, the evolution of the density matrix is unitary and governed by von Neumann equation \cite{vN}
\begin{equation}
\label{eq:2}
\frac{d\rho}{dt}= -i[\mathrm{H},\rho]
\end{equation}
where $\rho=\left| \psi\right\rangle  \left\langle \psi \right|$.  The eigenvalues of $\rho$ are given by $p_{j} \geq 0$ such that $\Sigma_{j}p_{j}=1$ which gives $\Sigma_{j} \dot{p}_{j}=0$, and eigenvectors of $\rho$ are given by $\left| \psi_{j}\right\rangle  $ where $j$ is used to label all non-zero eigenstates of $\rho$. 

An open system, which is central to understanding real systems, is the one that interacts with an environment.  Although the system and the environment can be considered to evolve unitarily, the state of the system alone is mixed and the evolution is not unitary due to interactions with the environment.   In such cases the system is described by a reduced density matrix which evolves according to a completely positive and trace preserving (CPTP) map such that $\rho (t) = \mathcal{E}_{t} (\rho (0))$, where $\rho (t)$ and $\rho(0)$ are the final and initial states, respectively \cite{BP}.  Indeed one can consider the reduced density matrix as $\rho(t)=tr_{E}\left( \left| \Psi(t)\right\rangle  \left\langle  \Psi(t)\right| \right)$ where $\left| \Psi(t)\right\rangle $ is the state of total system (including the environment) and $tr_{E}(.)$ is the trace over the environment degrees of freedom.  The diagonalized reduced density matrix is then $\rho=\Sigma_{j}p_{j} \left| \psi_{j}\right\rangle  \left\langle  \psi_{j}\right|$.  
A weakly coupled system with a large environment is expressed as a Markovian, CPTP, for which the evolution of the reduced density matrix is given by Lindblad equation \cite{BP,Li,M}. 
\begin{equation}
\label{eq:3}
\frac{d \rho}{dt}= - i[\mathrm{H}, \rho]+\Sigma_{k} \left( L_{k} \rho L^{\dagger}_{k}-\frac{1}{2}\left\lbrace L^{\dagger}_{k} L_{k}, \rho \right\rbrace \right), 
\end{equation}
where $k$ is the number of all distinct Lindblad operators that describe different types of environmental couplings.
The first term on the RHS of Eq. (\ref{eq:3}) is the usual unitary evolution term where $\mathrm{H}$ is the Hamiltonian of the system, and the second term represents the effect of the environment on the system and is called Lindblad superoperator, with $L_{k}$ called Lindblad operators that represent the environment coupling with the system.  

In this paper, we will consider two cases for an open quantum system.  In the first case the generator of evolution for the system is derived using both the standard method where the time-ordered expansion is considered under a time-dependent Hamiltonian, and using the Taylor expansion.  For the second case, we have no specific information about the generator of evolution.  In both cases we look at the rate of the fluctuation growth of an observable $A$ and its upper bound.  The results indicate that the information about the specific dynamics of the system does not affect the bound on the variations of the uncertainty of observable $A$.  Rather, all one needs to control the dynamics of the uncertainty is the information about the evolution of the observable $A$, and the state that is expressed by the density matrix $\rho$.  For completeness, we also show how the upper bound that was previously found for closed systems in Ref. \cite{CRBC} can be recovered from the bound for open systems in Appendix A and Appendix B, followed by an example to verify that in Appendix C.     

The rest of the paper is organized as follows. In Section II, we will find the time derivative of an observable in terms of the reduced density operator using two methods: the standard method based on time-ordered expansion that is valid for any finite time evolution, and the Taylor expansion which can be used for infinitesimal time steps.  In Section III, we will use the time evolution operator derived in Section II to look at the relation between time derivatives of the observable's average value and uncertainty.  In Section IV, we will study the relation between the time derivatives of the observable's average value and uncertainty without considering any specific time evolution operator.  We will discuss two examples in section V.   Our concluding remarks will be in Section VI.

\section{Time Evolution Operator and $d\left\langle A\right\rangle /dt$} 
In this section we will find the time evolution operator in an open quantum system using both the standard time-ordered method which uses the Dyson series, and the Taylor expansion.  We will show that the same result is obtained from both methods.  
The Taylor expansion which is used for infinitesimal time steps, is mainly used in numerical quantum dynamics.  For example, Taylor expansion is used to find a quantum algorithm for simulating the time evolution under a time-dependent Hamiltonian \cite{Be}.  However, one can arrive at the same result using the standard time-ordered expansion (Dyson series) which is applicable to any finite-time evolution under a time-dependent Hamiltonian \cite{Lou}.  For example, in \cite{MC} the discrete energy/mass spectrum of the Einstein-Vaz shell is considered to be a quantum system with discrete energy levels.  The authors in \cite{MC} use the Dyson series to describe the time evolution under a time-dependent Hamiltonian.  In another application, the Dyson series is used to derive the S-matrix and show how perturbation theory and Feynman rules emerge \cite{PS}.  

We will first look at the time evolution operator using the Taylor expansion.  Note that this is a specific time evolution operator which allows using the Taylor expansion since we are considering a unitary evolution operator which evolves the system from an initial time $t$ to a final time $t+dt$ such that $dt$ is very small.

Consider an infinitesimal change in $\langle A\rangle$ in Eq. (\ref{eq:1.1}).  We have, 
\begin{equation}
\label{eq:4}
d\langle A\rangle=\Sigma_{j}\left( dp_{j}\right) \langle \psi_{j}\vert A\vert \psi_{j}\rangle + \Sigma_{j}p_{j} d \left(\langle\psi_{j} \vert A\vert \psi_{j}\rangle  \right) 
\end{equation}
where we have used $\psi_{j}(t)=\psi_{j}$ for simplicity.
The evolution operator for the system is a unitary operator $T$ which is possibly time dependent and evolves the orthonormal basis $\left| \psi_{j} \right\rangle$ while preserving the structure and can be written as
\begin{equation}
\label{eq:5}
\vert \psi_{j}(t+dt)\rangle =T(t,t+dt)\vert \psi_{j}(t)\rangle.
\end{equation}
Note that since $\rho$ is Hermitian at any time, it must have an orthonormal basis of eigenvectors at any instant.  Then, since only unitary transformations preserve orthonormality, there must be a unitary operator $T$, possibly nonstationary, such that $\left| \psi_{j}(t+dt)\right\rangle =T(t,t+dt)\left|\psi_{j}(t)\right\rangle $.  This means that while orthonormal basis of $\rho$ evolve unitarily, its eigenvalues $p_{j}(t)$ undergo a nonunitary evolution.  Furthermore, let the Hermitian operator $\Omega(t)$ be the generator of this evolution.  Such a unitary operator is not uniquely determined by the dynamics, and does not represent the physical evolution of the system in general.  Hence we shall call its generator $\Omega$, pseudo-Hamiltonian since it only evolves the eigenstates. 
Time derivative of the state will then be given by
\begin{equation}
\label{eq:5.1}
\frac{d}{dt}\vert\psi_{j}(t)\rangle=\frac{d T(t)}{dt}\vert \psi_{j}(0)\rangle.
\end{equation}
Since we know that 
\begin{equation}
\label{eq:5.2}
i \left(\frac{dT(t)}{dt} \right) \vert \psi_{j}(0)\rangle=\Omega (t) T(t) \vert \psi_{j}(0)\rangle 
\end{equation}
and this holds for any arbitrary $\vert \psi_{j}(0)\rangle$, it follows that
\begin{equation}
\label{eq:5.3}
i \frac{dT(t)}{dt}=\Omega (t)T(t),
\end{equation}
where $\Omega$ is Hermitian.  Eq. (\ref{eq:5.3}) resembles the time-dependent Schr\"odinger equation for the operator $T(t)$ that evolves the eigenvectors $\left| \psi_{j}\right\rangle $ with time.

To find $T$ using the Taylor expansion, we use the same approach that is discussed in \cite{LBCN}.  We start by considering the Schr\"odinger equation and then applying it to Eq. (\ref{eq:5.3}) to find the evolution operator.  Using the Schr\"odinger equation, we have
\begin{equation}
\label{eq:6}
i \frac{\partial }{\partial t}\vert \psi(t)\rangle =\mathrm{H}(t) \vert \psi(t)\rangle
\end{equation}
with $\left| \psi(t)\right\rangle $ such that
\begin{equation}
\label{eq:7}
\vert \psi(t)\rangle =U(t,t_{0})\vert \psi(t_{0})\rangle.
\end{equation}
Furthermore, $U^{\dagger}U=I$ 
where $U(t,t_{0})$ is the evolution operator for the initial state $\left| \psi(t_{0})\right\rangle $ at time $t_{0}$ .
Define $U(t,t_{0})$ such that 
\begin{equation}
\label{eq:8}
i \frac{\partial U(t,t_{0})}{\partial t}=\mathrm{H}(t)U(t,t_{0})
\end{equation}
with $U(t_{0},t_{0})=I$.
When $\mathrm{H}$ is time-independent, then $U(t,t_{0})=exp\left( -i \mathrm{H}(t-t_{0})\right) $ with $\hbar=1$.  If, instead, $\mathrm{H}$ is time-dependent, then $U(t,t_{0})$ does not have an explicit and compact expression since in general, for $t\neq t'$ we have $[\mathrm{H}(t),\mathrm{H}(t')] \neq 0$. 
To find the time evolution operator $U(t,t_{0})$, let us expand both $U(t,t_{0})$ and $\mathrm{H}(t)$ is a Taylor series given by  
\begin{equation}
\label{eq:9}
U(t,t_{0})= \Sigma_{k=0}^{\infty} \frac{(t-t_{0})^{k}}{k!}U^{(k)}(t_{0},t_{0}), \quad \mathrm{H}(t)=\Sigma_{k=0}^{\infty} \frac{(t-t_{0})^{k}}{k!}\mathrm{H}^{(k)}(t_{0}),
\end{equation}
where 
\begin{equation}
\label{eq:9.1}
U^{(0)}(t,t_{0})=I, \, U^{(k)}(t_{0},t_{0})=\left( \frac{d^{k}}{dt^{k}}U(t,t_{0})\right) _{t=t_{0}}, \quad \mathrm{H}^{(0)}(t_{0})\equiv \mathrm{H}(t_{0}), \,  \mathrm{H}^{(k)}(t_{0})= \left( \frac{d^{k}}{dt^{k}} \mathrm{H}(t)\right) _{t=t_{0}}. 
\end{equation}
Substituting Eq. (\ref{eq:9}) into $i\dot{U}(t,t_{0})=\mathrm{H}(t_{0})U(t,t_{0})$, taking the derivative with respect to $t$ on the lefthand side, and then comparing terms of equal powers of $\Delta t \stackrel{\text{def}}{=}t-t_{0}$, one obtains  
\begin{equation}
\label{eq:9.2}
iU^{(1)}= \mathrm{H}^{(0)}\equiv \mathrm{H}(t_{0}), \quad iU^{(2)} = \mathrm{H}^{(0)}U^{(1)}+ \mathrm{H}^{(1)}=-i \mathrm{H}^{2}(t_{0})+ \mathrm{H}^{(1)}.
\end{equation}
Substituting these results back into the Taylor expansion Eq. (\ref{eq:9})we obtain to second order
\begin{equation}
\label{eq:10}
U(t,t_{0})= I-i \mathrm{H}(t_{0})\Delta t+\left(- i \mathrm{H}^{(1)}(t_{0})+\left( -i \mathrm{H}(t_{0})\right) ^{2}\right) \frac{\Delta t^{2}}{2} + \mathcal O(\Delta t^{3}),
\end{equation}
which can approximately be written as 
\begin{equation}
\label{eq:10.1}
U(t,t_{0}) \simeq I-i \mathrm{H}(t_{0})\Delta t
\end{equation}
where, as mentioned earlier, $\hbar$ is set to equal $1$.     
This result is obtained by using the Taylor expansion for Eq. (\ref{eq:8}) to find operator $U(t,t_{0})$ in Eq. (\ref{eq:7}) as $U(t,t_{0})=I-i \mathrm{H}(t_{0})dt$, since time step is infinitesimal.  

Next, we will find the evolution operator using the standard time-ordered expansion under a time-dependent Hamiltonian.  This method is applicable to any finite time evolution.  Use the formal solution to Eq. (\ref{eq:8}), which is the time-ordered expansion given by \cite{NC} 
\begin{equation}
\begin{aligned}
\label{eq:10.2}
U(t,t_{0}) &= \mathcal{T} \left[ e^{-i} \int_{t_{0}}^{t} \mathrm{H}(t') \,dt' \right] \\ 
 &= I + \Sigma_{n=1}^{\infty} \left( \frac{1}{i}\right)^{n} \int_{t_{0}}^{t} \,dt_{1} \int_{t_{0}}^{t_{1}} \,dt_{2} \cdots \int_{t_{0}}^{t_{n-1}} \mathrm{H}(t_{1})\mathrm{H}(t_{2}) \cdots \mathrm{H}(t_{n}) \,dt_{n} \\ 
 &= I + \left( \frac{1}{i}\right) \int_{t_{0}}^{t} \mathrm{H}(t_{1}) \,dt_{1}+\left( \frac{1}{i}\right) ^{2} \int_{t_{0}}^{t} \,dt_{1} \int_{t_{0}}^{t_{1}} \mathrm{H}(t_{1}) \mathrm{H}(t_{2}) \,dt_{2}+ \cdots 
\end{aligned}
\end{equation}
where $\mathcal{T}$ is the time ordering operator, and $t_{1} \geq t_{2}\geq ... \geq t_{n}$.  Using Eq. (\ref{eq:10.2}), one can see that the 0th order term is the identity matrix $I$, as expected.  Moreover, we can see that 
\begin{equation}
\begin{aligned}
\label{eq:10.3}
\int_{t_{0}}^{t_{0}+\Delta t} \mathrm{H}(t_{1}) \,dt_{1}  &\approx \int_{t_{0}}^{t_{0}+\Delta t} \left( \mathrm{H}(t_{0})+\dot{\mathrm{H}}(t_{0})(t_{1}-t_{0}) + \cdots \right)\,dt_{1} \\
 &= \mathrm{H}(t_{0})+\dot{\mathrm{H}}(t_{0})\int_{0}^{\Delta t}\tau \,d \tau \\
 &= \mathrm{H}(t_{0})+\dot{\mathrm{H}}(t_{0})\frac{\Delta t^{2}}{2}
\end{aligned}
\end{equation}
where we have set $\tau=t_{1}-t_{0}$.  Multiplying by $-i$ with $\hbar=1$ yields the second and the third terms on the RHS of Eq. (\ref{eq:10}).    

Now, considering Eq. (\ref{eq:5.3}), and following the same steps taken in Eqs. (\ref{eq:9}, \ref{eq:10}), or alternatively, the steps taken in Eqs. (\ref{eq:10.2}, \ref{eq:10.3}), one can obtain a term for operator $T(t+dt)$ in Eq.(\ref{eq:5}) as
\begin{equation}
\label{eq:11}
T(t,t+dt) \simeq I-i\Omega (t)dt, 
\end{equation}
where $\Omega$ is Hermitian and describes the evolution of the eigenvectors of the system's density matrix $\rho(t)$ in open quantum systems.  Since $\Omega$ which governs the dynamics of the system's state in the presence of environmental interactions through evolving only the eigenstates is analogous to the system's Hamiltonian, we call it pseudo-Hamiltonian.  Therefore, it is not the same as the generalized Hamiltonian in the Lindblad master equation given by Eq. (\ref{eq:3}).  
Using the result found in Eq. (\ref{eq:11}), we can consider small changes in $\left|\psi \right\rangle $ that can be written as

\begin{equation}
\begin{aligned}
\vert \psi_{j}(t+dt)\rangle -\vert \psi_{j}(t)\rangle &= T(t,t+dt)\vert \psi_{j}(t)\rangle-\vert \psi_{j}(t)\rangle \\
d \vert \psi_{j}(t)\rangle &= \left( I-i\Omega(t) dt\right) \vert \psi_{j}(t)\rangle -\vert \psi_{j}(t)\rangle \\
 &= -i \Omega(t) \vert \psi_{j}(t)\rangle dt.
\end{aligned}
\label{eq:12}
\end{equation}
Using the result from Eq. (\ref{eq:12}), we can write
\begin{equation}
\label{eq:13}
d\langle \psi_{j}(t) \vert=i\langle \psi_{j}(t) \vert \Omega^{\dagger}(t) dt=i\langle \psi_{j}(t) \vert \Omega(t) dt
\end{equation}
since $\Omega$ is Hermitian.
We can now use the result in Eq. (\ref{eq:4}) to find the change in the average value of the observable $A$.  Substitute Eq. (\ref{eq:13}) into Eq.(\ref{eq:4}) to get
\begin{equation}
\label{eq:14}
d \langle A \rangle= \Sigma_{j}dp_{j}\langle\psi_{j} \vert A \vert \psi_{j}\rangle + \Sigma_{j}p_{j} \langle \psi_{j} \vert dA \vert \psi_{j} \rangle + \Sigma_{j} p_{j} \left( -iA\Omega \vert \psi_{j}\rangle dt + i\langle \psi_{j} \vert \Omega A dt\right).  
\end{equation}
Eq. (\ref{eq:14}) can be rewritten as
\begin{equation}
\label{eq:15}
d \langle A \rangle= \Sigma_{j}dp_{j}\langle\psi_{j} \vert A \vert \psi_{j}\rangle + \Sigma_{j}p_{j} \langle \psi_{j} \vert dA \vert \psi_{j} \rangle + \Sigma_{j} ip_{j} \langle \psi_{j} \vert \left[ \Omega, A\right]  \vert \psi_{j} \rangle dt. 
\end{equation}
Dividing Eq. (\ref{eq:15}) by $dt$ gives
\begin{equation}
\label{eq:16}
\frac{d\langle A\rangle}{dt}= \Sigma_{j}\frac{dp_{j}}{dt}\langle \psi_{j} \vert A \vert \psi_{j} \rangle + \Sigma_{j} p_{j} \left\langle  \psi_{j} \left| \partial_{t}A \right| \psi_{j} \right\rangle + i\Sigma_{j} p_{j} \langle \psi_{j} \vert \left[ \Omega, A\right] \vert \psi_{j} \rangle,
\end{equation}
that is,
\begin{equation}
\label{eq:17}
\frac{d\langle A\rangle}{dt}= \Sigma_{j}\frac{dp_{j}}{dt}\langle \psi_{j} \vert A \vert \psi_{j} \rangle + \left\langle  \partial_{t} A \right\rangle  + i\langle \left[ \Omega,A\right]  \rangle,
\end{equation}
where we have used $\frac{\partial A}{\partial t}=\partial_{t}A$.  Note that Eq. (\ref{eq:17}) is the time derivative of the average value $\left\langle A\right\rangle $ which, in general is given by
\begin{equation}
\begin{aligned}
\frac{d\langle A\rangle}{dt} &= \frac{d}{dt}tr\left( \rho A\right) &= tr\left( \dot{\rho}A\right) + tr\left( \rho \partial_{t}{A}\right)
 &= tr\left( \dot{\rho}A\right) + \left\langle \partial_{t}{A}\right\rangle. 
\end{aligned}
\label{eq:17.1} 
\end{equation}
It is important to clarify that here $\partial{A}/\partial t \neq dA/dt$ since this only captures the time dependence of the observable, independent of the dynamics of the state that is included in $\rho$ or the Hamiltonian.  To avoid confusions, we will keep the partial derivatives instead of setting them equal to $\dot{A}$.

Knowing that $\left| \psi_{j}(t)\right\rangle $ evolves unitarily under the Hermitian operator $\Omega$, one can write  
\begin{equation}
\begin{aligned}
\dot{\rho} &= \Sigma_{j} \dot{p}_{j}\left| \psi_{j}\right\rangle  \left\langle \psi_{j}\right| + \Sigma_{j} p_{j}\left| \dot{\psi}_{j}\right\rangle  \left\langle \psi_{j}\right| + \Sigma_{j} p_{j}\left| \psi_{j}\right\rangle  \left\langle \dot{\psi}_{j}\right|, \\
&= \Sigma_{j} \dot{p}_{j}\left| \psi_{j} \right\rangle \left\langle \psi_{j}\right| + \Sigma_{j} p_{j}(-i \Omega\left| \psi_{j}\right\rangle)\left\langle \psi_{j}\right| + \Sigma_{j} p_{j} \left\langle \psi_{j} \right| (i \left\langle \psi_{j}\right| \Omega),\\
&= \Sigma_{j} \dot{p}_{j}\left| \psi_{j}\right\rangle  \left\langle \psi_{j}\right| + i\left[ \rho,\Omega\right] . 
\end{aligned}
\label{eq:17.2}
\end{equation}
Using the result found in Eq. (\ref{eq:17.2}), we have 
\begin{equation}
\begin{aligned}
tr(\dot{\rho}A) &= tr\left( (\Sigma_{i} \dot{p}_{i}\left| \psi_{i}\right\rangle  \left\langle \psi_{i}\right| + i\left[ \rho,\Omega\right])A\right), \\ 
&= \Sigma_{i} \dot{p}_{i}\left\langle \psi_{i}\right| A\left| \psi_{i}\right\rangle + i tr\left( \left[ \rho,\Omega\right]A \right), \\
&=  \Sigma_{i} \dot{p}_{i}\left\langle \psi_{i}\right| A\left| \psi_{i}\right\rangle + itr\left( \rho \left[ \Omega,A \right]\right), \\
&= \Sigma_{i} \dot{p}_{i}\left\langle \psi_{i}\right| A\left| \psi_{i}\right\rangle + i\left\langle \left[ \Omega,A\right] \right\rangle,   
\end{aligned}
\label{eq:17.3}
\end{equation}
where in going from the second to the third line, we have used the cyclic property of the trace, ($tr(A[B,C])=tr(B[C,A])=tr(C[A,B])$).  Similarly, we have 
\begin{equation}
\label{eq:17.4}
tr(\dot{\rho}A^{2})=\Sigma_{j} \dot{p}_{j} \left\langle \psi_{j}\left| A^{2}\right| \psi_{j}\right\rangle  + i\left\langle \left[ \Omega, A^{2}\right] \right\rangle. 
\end{equation}

Using the result found in Eq. (\ref{eq:17.3}), and comparing Eq. (\ref{eq:17}) with Eq. (\ref{eq:17.1}), one can realize that the first and the third terms on the RHS of both equations represent $tr(\dot{\rho}A)$, while the second term on the RHS of Eq. (\ref{eq:17}) indicates that $tr(\rho \partial_{t}A)= \left\langle \partial_{t}A\right\rangle $.

We can now use the results found for the average value of $A$ given by Eq. (\ref{eq:1.1}) and its derivative given by Eq. (\ref{eq:17}), to find the uncertainty $\sigma_{A}$ and its time derivative $\dot{\sigma}_{A}$ to obtain a bound on the rate of fluctuation growth.

\section{Fluctuation Growth with Knowledge of the pseudo-Hamiltonian}
We can now find the rate of fluctuation growth of observable $A$ in an open system under the time evolution operator $T(t)$ that we found in Section I.  To find the dynamics of fluctuations of a time-dependent observable $A$, we start by using its variance that is given by 
\begin{equation}
\label{eq:20}
\sigma_{A}^{2}= \langle \Delta A^{2}\rangle=\Sigma_{j} p_{j}\langle \psi_{j} \vert A^{2}\vert \psi_{j}\rangle - \left( \Sigma_{j} p_{j}\langle \psi_{j} \vert A\vert \psi_{j}\rangle \right) ^{2}
\end{equation}
where $\Delta A \stackrel{\text{def}}{=} A-\left\langle A\right\rangle $ which gives $\sigma^{2}_{A}=\left\langle A^{2}\right\rangle -\left\langle A\right\rangle ^{2}$.
Taking the time derivative of $\sigma_{A}$ yields
\begin{equation}
\label{eq:21}
\frac{d}{dt}\left( \sigma_{A}\right) =\frac{1}{2\sigma_{A}}\left[ \frac{d\left\langle A^{2}\right\rangle }{dt}-2\left\langle A\right\rangle \frac{d\left\langle A\right\rangle }{dt}\right]. 
\end{equation}
Using Eq. (\ref{eq:17}), the term $d\left\langle A\right\rangle /dt$ in Eq. (\ref{eq:21}) becomes
\begin{equation}
\label{eq:22} 
\frac{d\left\langle A\right\rangle }{dt}=\Sigma_{j}\dot{p}_{j}\left\langle \psi_{j}\vert A\vert \psi_{j}\right\rangle +\left\langle \partial_{t}A\right\rangle +i\Sigma_{j}p_{j}\left\langle \psi_{j} \vert \left[ \Omega,A\right] \vert \psi_{j} \right\rangle  .
\end{equation}
Similarly, time derivative of $\left\langle A^{2}\right\rangle $ in Eq. (\ref{eq:21}) is given by
\begin{equation}
\label{eq:23} 
\frac{d\left\langle A^{2}\right\rangle }{dt}=\Sigma_{j}\dot{p}_{j}\left\langle \psi_{j}\vert A^{2}\vert \psi_{j}\right\rangle +\left\langle \partial_{t} A^{2}\right\rangle +i\Sigma_{j}p_{j}\left\langle \psi_{j} \vert \left[ \Omega,A^{2}\right] \vert \psi_{j} \right\rangle ,
\end{equation}
where $\dot{p}_{j} \stackrel{\text{def}}{=}dp_{j}/dt$.
Substituting Eqs. (\ref{eq:22}) and (\ref{eq:23}) into Eq. (\ref{eq:21}) and squaring it, we obtain

\begin{multline}
\label{eq:24}
\left( \frac{d \sigma_{A}}{dt}\right) ^{2}=\frac{1}{4\sigma^{2}_{A}} \bigg[  \Sigma_{j}\dot{p}_{j}\left\langle \psi_{j}\vert A^{2} \vert\psi_{j} \right\rangle +\left\langle \partial_{t} A^{2} \right\rangle +i\Sigma_{j}p_{j}\left\langle \psi_{j}\vert \left[ \Omega, A^{2}\right] \vert\psi_{j}\right\rangle \\
-2\Sigma_{j}p_{j}\left\langle \psi_{j}\vert A \vert\psi_{j}\right\rangle \left( \Sigma_{j}\dot{p}_{j}\left\langle \psi_{j}\vert A \vert\psi_{j}\right\rangle +\left\langle \partial_{t} A \right\rangle +i\Sigma_{j}p_{j}\left\langle \psi_{j}\vert\left[ \Omega,A\right] \vert\psi_{j}\right\rangle \right) \bigg]  ^{2}.
\end{multline}
Eq. (\ref{eq:24}) can be rewritten as
\begin{multline}
\label{eq:25}
\left( \frac{d \sigma_{A}}{dt}\right) ^{2}=\frac{1}{4\sigma^{2}_{A}} \left( \Sigma_{j} \dot{p}_{j}\left\langle \psi_{j} \vert A^{2} \vert \psi_{j}\right\rangle +\left\langle \partial_{t} A^{2}\right\rangle +i\left\langle \left[ \Omega, A^{2}\right] \right\rangle \right. \\
\left. -2\left\langle A\right\rangle \Sigma_{j} \dot{p}_{j}\left\langle \psi_{j} \vert A\vert \psi_{j}\right\rangle -2\left\langle A\right\rangle \left\langle \partial_{t} A\right\rangle -2i\left\langle A\right\rangle \left\langle \left[ \Omega, A\right] \right\rangle \right)  ^{2}.   
\end{multline}
To simplify Eq. (\ref{eq:25}), use the results found in Eqs. (\ref{eq:17.3}) and (\ref{eq:17.4}) to write
\begin{equation}
\label{eq:25.1}
\left( \frac{d\sigma_{A}}{dt}\right) ^{2} = \frac{1}{4\sigma^{2}_{A}}\left( tr\left( \dot{\rho}A^{2}\right) - 2\left\langle A\right\rangle tr\left( \dot{\rho}A\right) + \left\langle \partial_{t}A^{2}\right\rangle - 2\left\langle A\right\rangle \left\langle \partial_{t}A\right\rangle \right) ^{2}.
\end{equation}
We also know that 
\begin{equation}
\label{eq:26}
tr\left( \dot{\rho}A^{2}\right)  - 2\left\langle A\right\rangle tr(\dot{\rho}A) = tr\left( \dot{\rho}\Delta A^{2}\right) 
\end{equation}
where we have used $\Delta A ^{2}=(A-\left\langle A\right\rangle )^{2}$ and multiplied it by $\dot{\rho}$ and finally, performed the trace to get
$tr(\dot{\rho} \Delta A^{2})= tr(\dot{\rho} A^{2})-2<A> tr(\dot{\rho}A)+ <A>^{2}tr(\dot{\rho})$ since $tr(\dot{\rho})=\Sigma_{j}\dot{p}_{j}=0$.  
Using Eq. (\ref{eq:26}), Eq. (\ref{eq:25.1}) yields
\begin{equation}
\label{eq:27}
\left( \frac{d\sigma_{A}}{dt}\right) ^{2} = \frac{1}{4 \sigma^{2}_{A}}\left( tr\left(\dot{\rho}\Delta A^{2}\right)  + \left\langle \partial_{t}A^{2}\right\rangle - 2\left\langle A\right\rangle \left\langle \partial_{t}A\right\rangle \right) ^{2}.
\end{equation}
Then using $2 Cov(X,\dot{X})\stackrel{\text{def}}{=}\left\langle \left\lbrace X, \dot{X}\right\rbrace \right\rangle -2\left\langle X\right\rangle \left\langle \dot{X}\right\rangle $,  we can rewrite Eq. (\ref{eq:27}) as
\begin{equation}
\label{eq:28}
\left( \frac{d\sigma_{A}}{dt}\right) ^{2} = \frac{1}{4\sigma^{2}_{A}}\left( tr\left(\dot{\rho}\Delta A^{2}\right)  + 2 Cov(A,\partial_{t}A) \right) ^{2},
\end{equation}
where $Cov(X, \dot{X})$ measures the degree to which the fluctuations of two observables $X$ and $\dot{X}$ in a density matrix are correlated.
Note that since the term $tr\left( \dot{\rho}\Delta A^{2}\right) =tr\left( \dot{\rho} A^{2}\right) -2\left\langle A\right\rangle tr \left( \dot{\rho} A\right) $ only depends on the time derivative of $\rho$ and not the time derivative of the observable $A$ and since the non-unitary evolution is included in the eigenvalue of $\rho$, this term captures only the non-unitary contribution to the change in uncertainty.  However, the term $\frac{d (\sigma^{2}_{A})}{dt}=\frac{d\left\langle \Delta A^{2}\right\rangle }{dt}= \frac{d(tr(\rho (\Delta A)^{2})}{dt}$ captures both non-unitary and time-dependence of the observable.  Therefore, we generally have $tr\left( \dot{\rho}\Delta A^{2}\right) < \frac{d \sigma^{2}_{A}}{dt}$.

As a side remark, note that we can use $\frac{d (\sigma^{2}_{A})}{dt}=\frac{d}{dt} \left\langle \Delta A^{2}\right\rangle$, which we know is always true, and rewrite it in terms of trace to get  
\begin{equation}
\label{eq:35}
\frac{d (\sigma^{2}_{A})}{dt}=\frac{d}{dt} \left\langle \Delta A^{2}\right\rangle = \frac{d}{dt} tr\left( \rho (\Delta A)^{2}\right) = tr\left( \dot{\rho}(\Delta A)^{2}\right) + tr\left( \rho \frac{d(\Delta A^{2})}{dt} \right) . 
\end{equation}
Then using Eq. (\ref{eq:35}), we can rewrite  Eq. (\ref{eq:28}) as
\begin{equation}
\label{eq:36}
tr\left( \dot{\rho}\Delta A^{2}\right) +2 Cov \left( A,\partial_{t} A\right) = \frac{d(\sigma^{2}_{A})}{dt},
\end{equation}
that is, $(d\sigma_{A}/dt)^{2}=(1/4\sigma^{2}_{A})(d(\sigma^{2}_{A})/dt)$.

To find how the fluctuation growth of observable $A$ is bounded, we start by using the Cauchy-Schwarz inequality written as

\begin{equation}
\label{eq:38}
\vert tr\left( \rho (\Delta A) (\partial_{t}A) \right) \vert ^{2} \leq tr\left( \rho \Delta A^{2}\right) tr\left( \rho (\partial_{t}A)^{2}\right). 
\end{equation}
Since we have that $Cov (A,\partial_{t}A)= tr\left( \rho (\Delta A) \partial_{t}A \right) $, we can write from Eq. (\ref{eq:38})
\begin{equation}
\label{eq:39}
\vert Cov(A,\partial_{t}A)\vert ^{2} \leq \left\langle \Delta A^{2}\right\rangle \left\langle (\partial_{t}A)^{2}\right\rangle ,
\end{equation}
or, alternatively,
\begin{equation}
\label{eq:40}
\vert Cov(A,\partial_{t}A)\vert ^{2} \leq \sigma^{2}_{A} \left\langle (\partial_{t}A)^{2}\right\rangle . 
\end{equation}
Given that for any real $x$ and $y$ we have   
\begin{equation}
\label{eq:41}
(x+y)^{2} \leq 2(x^{2}+y^{2}),
\end{equation}
we can write
\begin{equation}
\label{eq:42}
\left( tr\left( \dot{\rho}\Delta A^{2}\right) + 2Cov(A, \partial_{t}A)\right) ^{2} \leq 2\left( \left( tr(\dot{\rho}\Delta A^{2})\right) ^{2} + 4 \left( Cov(A,\partial_{t}A)\right) ^{2}\right) .
\end{equation}
Using Eq. (\ref{eq:36}) to write Eq. (\ref{eq:42}) yields
\begin{equation}
\label{eq:43}
\left( \frac{d \sigma^{2}_{A}}{dt}\right) ^{2} \leq 2\left( tr(\dot{\rho}\Delta A^{2})\right) ^{2} + 8\left( Cov(A,\partial_{t}A)\right) ^{2}
\end{equation}
which can be written as
\begin{equation}
\label{eq:44}
4\sigma^{2}_{A} \dot{\sigma}^{2}_{A} \leq 2\left( tr(\dot{\rho}\Delta A^{2})\right) ^{2} + 8\left( Cov(A,\partial_{t}A)\right) ^{2}.
\end{equation}
Eq. (\ref{eq:44}) can be rearranged to give
\begin{equation}
\label{eq:45}
\frac{1}{2}\sigma^{2}_{A} \dot{\sigma}^{2}_{A} -\frac{1}{4} \left( tr(\dot{\rho}\Delta A^{2})\right) ^{2} \leq \left( Cov(A,\partial_{t}A)\right) ^{2}.
\end{equation}
We can now use Eq. (\ref{eq:40}) to write Eq. (\ref{eq:45}) as
\begin{equation}
\label{eq:45.1}
\frac{1}{2}\sigma^{2}_{A} \dot{\sigma}^{2}_{A} -\frac{1}{4} \left( tr(\dot{\rho}\Delta A^{2})\right) ^{2} \leq \sigma^{2}_{A} \left\langle (\partial_{t}A)^{2}\right\rangle ,
\end{equation}
which can be rewritten as
\begin{equation}
\label{eq:46}
\dot{\sigma}^{2}_{A} \leq 2 \left( \left\langle (\partial_{t}A)^{2}\right\rangle +\frac{(tr(\dot{\rho}\Delta A^{2}))^{2}}{4\sigma^{2}_{A}}\right) .
\end{equation}

This inequality expresses that the rate of fluctuation growth of observable $A$ is upper bounded by the terms on the RHS of Eq. (\ref{eq:46}).  The first term on the RHS captures how much the observable itself is changing with time which can be due to externally-induced changes, and the second term on the RHS expresses the change of uncertainty due to the state's change plus the effect of fluctuations correlated with its motion due to its generator, which is $\Omega$ here.  Thus the effect of $\Omega$ is included in this term.  Note that this result is found based on a specific pseudo-Hamiltonian $\Omega$ that was found when we used both the time-ordered expansion as well as the Taylor expansion to find the evolution operator $T(t)$ based on the assumption that the evolution operator remains unitary when $dt$ remains very small.  Alternatively, we can consider an open system without having any information about its time evolution operator.  In the next section we will consider such a system and see how fluctuations of its uncertainty are related with the average value of its time derivative.

\section{Fluctuation Growth without Knowledge of the pseudo-Hamiltonian}
In this section we will consider an open system without any knowledge about the generator of evolution.  In this case, all we have is the evolution of the observable $A$ and its uncertainty, which means we only know that $\Delta A= A-\left\langle A\right\rangle$ and that $\left\langle \Delta A^{2}\right\rangle = \sigma^{2}_{A}=\left\langle A^{2}\right\rangle -\left\langle A\right\rangle ^{2}$.  So, we have 
\begin{equation}
\label{eq:50}
\frac{d \sigma^{2}_{A}}{dt}=\frac{d}{dt}tr\left( \rho(\Delta A^{2})\right) = tr\left( \dot{\rho}(\Delta A^{2})\right) + tr\left( \rho \frac{d}{dt}(\Delta A^{2})\right) . 
\end{equation}
For the second term on the RHS of Eq. (\ref{eq:50}), we have 
\begin{equation} 
\label{eq:51}
\begin{split}
\frac{d}{dt}(\Delta A^{2}) & = \frac{d}{dt}\left( A^{2}-2A\left\langle A\right\rangle +\left\langle A\right\rangle ^{2}\right) \\
 & = (\partial_{t}{A})A+A\partial_{t}{A}-2\partial_{t}{A}\left\langle A\right\rangle -2A\frac{d\left\langle A\right\rangle }{dt}+2\left\langle A\right\rangle \frac{d\left\langle A\right\rangle }{dt}.
\end{split}
\end{equation} 
Therefore, $tr\left(  \rho \frac{d}{dt} (\Delta A^{2})\right) $ reduces to 
\begin{equation} 
\label{eq:52}
\begin{split}
tr\left(  \rho \frac{d}{dt} (\Delta A^{2})\right)  & = tr\left( \rho (\partial_{t}{A}) A\right) +tr\left( \rho A \partial_{t}{A}\right) -2\left\langle A\right\rangle tr\left( \rho \partial_{t}{A}\right) -2\left\langle A\right\rangle tr\left( \rho \partial_{t}{A}\right) +2\left\langle A\right\rangle tr\left( \rho \partial_{t}{A}\right) \\
  & = tr\left( \rho (\partial_{t}{A}) A\right) +tr\left( \rho A \partial_{t}{A}\right) -2\left\langle A\right\rangle tr\left( \rho \partial_{t}{A}\right) \\
  & = 2Re\left\langle A\partial_{t}{A}\right\rangle -2\left\langle \partial_{t}{A}\right\rangle \left\langle A\right\rangle ,
\end{split}
\end{equation}
where $tr(\rho \partial_{t}{A}) = \left\langle \partial_{t}{A}\right\rangle $.  Also, since $A$ is Hermitian we have $tr\left( \rho A^{\dagger}\right) =tr\left( \rho A\right) = (tr\left( \rho A\right) )^{*}$ where in the last equality we have used the fact that $\left\langle A\right\rangle \in \mathbb{R}$.  These results  can be used to write 
\begin{equation}
\label{eq:52.1}
\begin{split}
\left( tr\left( \rho A \partial_{t}{A}\right) \right) ^{*} = tr\left( (\rho A \partial_{t}{A})^{\dagger}\right)  =tr\left( (\partial_{t}{A})^{\dagger} A \rho\right)  = tr \left(  \rho \left( A \partial_{t}{A}\right)^{\dagger}\right), 
\end{split}
\end{equation}
that is, 
\begin{equation}
\label{eq:52.2}
 \left( tr\left( \rho A \partial_{t}{A}\right)\right)  ^{*} = tr \left( \rho \left( A \partial_{t}{A}\right)^{\dagger}\right).  
\end{equation}
Eq. (\ref{eq:52.2}) yields,
\begin{equation}
\label{eq:52.3}
tr(\rho A \partial_{t}{A})+\left( tr(\rho A \partial_{t}{A})\right) ^{*}=2 Re\left( tr\left( \rho A \partial_{t}{A}\right) \right) .
\end{equation}
We also know that
\begin{equation}
\label{eq:53}
2Re\left\langle A\partial_{t}{A}\right\rangle -2\left\langle \partial_{t}{A}\right\rangle \left\langle A\right\rangle = 2 Re Cov(A,\partial_{t}{A})
\end{equation}
given that,
\begin{equation}
\label{eq:54.1}
Re \left( Cov(A, \partial_{t}{A})\right) = Re\left( \left\langle A\partial_{t}{A}\right\rangle -\left\langle A\right\rangle \left\langle \partial_{t}{A}\right\rangle \right) = Re\left\langle A \partial_{t}{A}\right\rangle -\left\langle A\right\rangle Re\left\langle \partial_{t}{A}\right\rangle = Re\left\langle A \partial_{t}{A}\right\rangle -\left\langle A\right\rangle \left\langle \partial_{t}{A}\right\rangle , 
\end{equation}
since $\left\langle \partial_{t}{A}\right\rangle $ is real ($(\partial_{t}{A})^{\dagger}=\left( \partial{A}/\partial{t}\right) ^{\dagger}=d A^{\dagger}/dt=\partial_{t}{A}$).  Note that $d(\Delta A^{2})/dt$ in the previous case where the generator of evolution was known, was given by Eq. (\ref{eq:36}). 
We also know that in general
\begin{equation}
\label{eq:54}
\begin{split}
tr\left( \rho \Delta A (\partial_{t}{A})\right) & = tr\left( \rho (A-\left\langle A\right\rangle )\partial_{t}{A}\right) = tr \left( \rho A \partial_{t}{A}\right) -\left\langle A\right\rangle tr\left( \rho \partial_{t}{A}\right) \\
 & = \left\langle A \partial_{t}{A}\right\rangle -\left\langle A\right\rangle \left\langle \partial_{t}{A}\right\rangle  = Cov(A, \partial_{t}{A}).
 \end{split}  
\end{equation}

Next, use of the Cauchy-Schwarz inequality
\begin{equation}
\label{eq:55}
\vert tr\left( \rho \Delta A (\partial_{t}{A})\right) \vert ^{2} \leq tr\left( \rho \Delta A^{2}\right) tr\left( \rho (\partial_{t}{A})^{2}\right) 
\end{equation}
together with Eq. (\ref{eq:54}) gives 
\begin{equation}
\label{eq:56}
\vert Cov(A,\partial_{t}{A})\vert ^{2} \leq \sigma^{2}_{A} \left\langle (\partial_{t}{A})^{2}\right\rangle ,  
\end{equation}
where we have used $tr\left( \rho \Delta A^{2}\right) = \left\langle \Delta A^{2}\right\rangle =\sigma^{2}_{A}$.  
Using Eqs. (\ref{eq:50}, \ref{eq:53}), we can write 
\begin{equation}
\label{eq:57}
\left( \frac{d \sigma^{2}_{A}}{dt}\right) ^{2}=\left( tr\left( \dot{\rho} \Delta A^{2}\right) +2Re Cov(A,\partial_{t}{A})\right) ^{2}.
\end{equation}
Exploiting the inequality 
\begin{equation}
\label{eq:58}
(x+y)^{2} \leq 2(x^{2}+y^{2}),
\end{equation}
valid for all real $x$ and $y$, we can write
\begin{equation}
\label{eq:59}
\left( tr\left( \dot{\rho} \Delta A^{2}\right) +2Re Cov(A,\partial_{t}{A})\right) ^{2} \leq 2\left( (tr\left( \dot{\rho} \Delta A^{2}\right))^{2} +4(Re Cov(A,\partial_{t}{A}))^{2}\right) . 
\end{equation}
From Eqs. (\ref{eq:57}) and (\ref{eq:59}), we get
\begin{equation}
\label{eq:60}
\frac{1}{2}\left( \frac{d \sigma^{2}_{A}}{dt}\right) ^{2} - (tr\left( \dot{\rho} \Delta A^{2}\right))^{2} \leq 4 \left( Re Cov(A,\partial_{t}{A})\right) ^{2} \leq 4\vert Cov(A,\partial_{t}{A}) \vert^{2} .
\end{equation}
Eq. (\ref{eq:60}) can be rewritten as
\begin{equation}
\label{eq:61}
\frac{1}{2}\left( 4 \sigma^{2}_{A} \dot{\sigma}^{2}_{A}\right)  \leq 4 \sigma^{2}_{A} \left\langle (\partial_{t}{A})^{2}\right\rangle +(tr\left( \dot{\rho} \Delta A^{2}\right))^{2}, 
\end{equation}
or, alternatively, 
\begin{equation}
\label{eq:62}
\dot{\sigma}^{2}_{A} \leq 2 \left( \left\langle (\partial_{t}{A})^{2}\right\rangle  +\frac{(tr\left( \dot{\rho} \Delta A^{2}\right))^{2}}{4\sigma^{2}_{A}}\right).
\end{equation}

The inequality in Eq. (\ref{eq:62}) gives an upper bound on the rate of fluctuations of observable $A$, which is expressed by $\dot{\sigma}^{2}_{A}$.  
While the first term on the RHS indicates how fast the observable is changing with time, the second term on the RHS analyzes the rate of change of uncertainty of observable $A$.  Although Eqs. (\ref{eq:46}, \ref{eq:62}) are the same, unlike the terms on the RHS of Eq. (\ref{eq:46}), the terms on the RHS of Eq. (\ref{eq:62}) are not limited to merely environmental influences on the spread of the observable since they are not derived based on assuming a specific type of evolution. Rather, in Eq. (\ref{eq:62}) $d \rho/dt$ could evolve stricly unitarily via $\mathrm{H}$ (in a closed system) as given by Eq. (\ref{eq:2}), or it could evolve by a master equation for the reduced system density matrix $\rho$ in an open system which contains both $\mathrm{H}$ and environmental Lindblad operators $L$, given by Eq. (\ref{eq:3}).  Moreover, the solution $\rho(t)$, that is invloved in the first term on the RHS of Eq. (\ref{eq:62}), may also come from either Eq. (\ref{eq:2}) or Eq. (\ref{eq:3}).

Alternatively, one could derive Eq. (\ref{eq:62}) by considering the symmetrized version of the covariance given by
\begin{equation}
\label{eq:P4.1}
Cov(A,\partial_{t}A)=\frac{1}{2}(\left\langle A\partial_{t}A+(\partial_{t}A)A\right\rangle -\left\langle A\right\rangle \left\langle \partial_{t}A\right\rangle ).
\end{equation}
Then, one can write Eq. (\ref{eq:52}) as
\begin{equation}
\label{eq:P4.2}
tr\left(  \rho \frac{d}{dt} (\Delta A^{2})\right) = tr\left( \rho (\partial_{t}{A}) A\right) +tr\left( \rho A \partial_{t}{A}\right) -2\left\langle A\right\rangle tr\left( \rho \partial_{t}{A}\right) = 2Cov(A,\partial_{t}A),
\end{equation}
which gives Eq. (\ref{eq:51}) as
\begin{equation}
\label{eq:P4.3}
\frac{d}{dt}\sigma_{A}=\frac{1}{2\sigma_{A}}\left( tr(\dot{\rho}\Delta A^{2})+2Cov(A,\partial_{t}A)\right) .
\end{equation}
Using Eq. (\ref{eq:58}), gives
\begin{equation}
\label{eq:P4.4}
\left( 2Cov(A,\partial_{t}A)\right) ^{2} \leq 2\left( tr(\rho (\Delta A)\dot{A})^{2}+tr(\rho \dot{A}\Delta A)^{2}\right), 
\end{equation}
and using Cauchy-Schwarz on the RHS of Eq. (\ref{eq:P4.4}) leads to
\begin{equation}
\label{eq:P4.5}
\left( 2Cov(A,\partial_{t}A)\right) ^{2} \leq 2\left( tr\left( \rho\Delta A^{2}\right) tr\left( \rho (\partial_{t}A)^{2}\right) + tr\left( \rho (\partial_{t}A)^{2}\right)tr\left( \rho \Delta A^{2}\right) \right), 
\end{equation}
that is
\begin{equation}
\label{eq:P4.6}
\left( Cov(A,\partial_{t}A)\right) ^2 \leq tr\left( \rho (\partial_{t}A)^{2}\right)tr\left( \rho \Delta A^{2}\right),
\end{equation}
or, simply
\begin{equation}
\label{eq:P4.7}
Cov(A,\partial_{t}A) \leq \sigma_{A} \sqrt{\left\langle \partial_{t}A\right\rangle },
\end{equation}
since the RHS of Eq. (\ref{eq:P4.6}) is always positive and Eq. (\ref{eq:P4.7}) always holds regardless of whether $Cov(A,\partial_{t}A)$ is positive or negative.  Using Eq. (\ref{eq:P4.7}), one can rewrite Eq. (\ref{eq:P4.3}) as
\begin{equation}
\label{eq:P4.8}
\dot{\sigma}_{A} \leq \frac{1}{2\sigma_{A}}\left( tr\left( \dot{\rho}\Delta A^{2}\right) +2\sigma_{A}\sqrt{\left\langle (\partial_{t}A)^{2}\right\rangle }\right) .
\end{equation}
Finally, squaring both sides of Eq. (\ref{eq:P4.8}) and using Eq. (\ref{eq:58}) gives
\begin{equation}
\label{eq:P4.9}
(\dot{\sigma}_{A})^{2} \leq 2 \left( \frac{(tr(\dot{\rho}\Delta A^{2})^{2}}{4\sigma^{2}_{A}} +\left\langle \partial_{t}A^{2}\right\rangle \right) , 
\end{equation}
which is the same result found in Eq. (\ref{eq:62}).

The fact that the two inequalities given by Eqs. (\ref{eq:46}) and (\ref{eq:62}) are exactly the same, suggests that the dynamics of the system does not affect the rate of change of variance of observable $A$.  This is due to the fact that the RHS of both equations only depend on the explicit time dependence of the observable and the time derivative of $\rho$. Due to this result, the inequality is applicable to situations where the open system is partially understood.  That is, all we need is $\dot{\rho}$ and the observable, even if the generator is not known.  In Appendix A we will show how the result found in Eq. (\ref{eq:62}) reduces to the one we arrived at in Ref. \cite{CRBC} for closed systems.  Furthermore, in Appendix B we will show how to recover the inequality for closed systems from Eq. (\ref{eq:P4.3}), followed by an illustrative example in Appendix C, which further verifies that the inequality given by Eq. (\ref{eq:62}) is the bound for open systems.

To verify the inequality we found in Sections III and IV, we will now look at two examples with and without a known Hamiltonian.

\section{Illustrative examples}

In the following section, we aim to verify the validity of our fluctuation
growth inequality for a straightforward open quantum system. In particular, we
examine a qubit that is influenced by amplitude damping noise
\cite{NC,laleh11,peter14}.

\subsection{Noise model}

This model of quantum noise illustrates energy relaxation phenomena, such as
spontaneous emission, wherein a two-level quantum system (i.e., a qubit)
dissipates energy (i.e., emits a photon) to its surroundings by transitioning
from the excited state $\left\vert 1\right\rangle $ to the ground state
$\left\vert 0\right\rangle $. This process is a non-unitary irreversible one
that can be mathematically represented through the amplitude damping channel%
\begin{equation}
\Lambda_{\mathrm{AD}}\left(  \rho\right)  \overset{\text{def}}{=}E_{0}\rho
E_{0}^{\dagger}+E_{1}\rho E_{1}^{\dagger}\text{,} \label{LAD}%
\end{equation}
where $\rho$ is the initial density matrix of the qubit. The terms $E_{0}$ and
$E_{1}$ are the so-called Kraus operators given by%
\begin{equation}
E_{0}\overset{\text{def}}{=}\left(
\begin{array}
[c]{cc}%
1 & 0\\
0 & \sqrt{1-\gamma}%
\end{array}
\right)  \text{, and }E_{1}\overset{\text{def}}{=}\left(
\begin{array}
[c]{cc}%
0 & \sqrt{\gamma}\\
0 & 0
\end{array}
\right)  \text{,}%
\end{equation}
respectively, with $E_{0}^{\dagger}E_{0}+E_{1}^{\dagger}E_{1}=\mathbf{1}$
(i.e., the channel is trace-preserving). Clearly $\mathbf{1}$ denotes here the
identity operator. The unitless parameter $\gamma\in\left[  0\text{,
}1\right]  $ represents the damping probability with $0\leq\gamma\leq1$,
indicating the likelihood of the qubit's relaxation. To interpret the effect
of this noise on the qubit, we note that if $\rho\overset{\text{def}}%
{=}\left\vert \psi\right\rangle \left\langle \psi\right\vert $ with
$\left\vert \psi\right\rangle \overset{\text{def}}{=}\alpha\left\vert
0\right\rangle +\beta\left\vert 1\right\rangle $ and $\left\vert
\alpha\right\vert ^{2}+\left\vert \beta\right\vert ^{2}=1$, the matrix
representation of $\Lambda_{\mathrm{AD}}\left(  \rho\right)  $ with respect to
the single-qubit computational basis becomes%
\begin{equation}
\Lambda_{\mathrm{AD}}\left(  \rho\right)  =\left(
\begin{array}
[c]{cc}%
\left\vert \alpha\right\vert ^{2}+\gamma\left\vert \beta\right\vert ^{2} &
\sqrt{1-\gamma}\alpha\beta^{\ast}\\
\sqrt{1-\gamma}\alpha^{\ast}\beta & \left(  1-\gamma\right)  \left\vert
\beta\right\vert ^{2}%
\end{array}
\right)  \text{.} \label{lad}%
\end{equation}
Therefore, from Eq. (\ref{lad}) we observe that: i) the ground state gains
probability weight since $\left\vert \alpha\right\vert ^{2}\rightarrow
\left\vert \alpha\right\vert ^{2}+\gamma\left\vert \beta\right\vert ^{2}$, the
population of the excited state decreases since $\left\vert \beta\right\vert
^{2}\rightarrow$ $\left(  1-\gamma\right)  \left\vert \beta\right\vert ^{2}$
and, finally, the off-diagonal terms (i.e., coherence terms) shrink since
$\alpha\beta^{\ast}\rightarrow\sqrt{1-\gamma}\alpha\beta^{\ast}\alpha
\beta^{\ast}$ and $\alpha^{\ast}\beta\rightarrow\sqrt{1-\gamma}\alpha^{\ast
}\beta$.

In the Born-Markov approximation in which one assumes a weak system-environment coupling and a short correlation time in the environment,
the amplitude damping channel corresponds to a Markov noise process. The continuous time dynamics of such a Markov noise process that represents the amplitude damping noise model is described by the Lindblad master equation
given by%
\begin{equation}
\frac{d\rho\left(  t\right)  }{dt}=-i\left[  \mathrm{H}\text{, }\rho\left(
t\right)  \right]  +\mathcal{D}\left[  \rho\left(  t\right)  \right]  \text{.}
\label{markov}%
\end{equation}
The term $-i\left[  \mathrm{H}\text{, }\rho\left(  t\right)  \right]  $ (with
$\hslash$ set equal to one) is the unitary part related to the
Schr\"{o}dinger evolution, while $\mathcal{D}\left[  \rho\left(  t\right)
\right]  $ in Eq. (\ref{markov}) is the so-called dissipator describing the
dissipative aspects of the dynamics. It captures the loss of energy due to
coupling with the environment, modeling spontaneous emission or relaxation
from $\left\vert 1\right\rangle $ to $\left\vert 0\right\rangle $. In the case
of amplitude damping, $\mathcal{D}\left[  \rho\left(  t\right)  \right]  $ is
defined as%
\begin{equation}
\mathcal{D}\left[  \rho\left(  t\right)  \right]  \overset{\text{def}}{=}%
L\rho\left(  t\right)  L^{\dagger}-\frac{1}{2}\left\{  L^{\dagger}L\text{,
}\rho\left(  t\right)  \right\}  =\Gamma\left[  \sigma_{-}\rho\left(
t\right)  \sigma_{+}-\frac{1}{2}\left\{  \sigma_{+}\sigma_{-}\text{, }%
\rho\left(  t\right)  \right\}  \right]  \text{,} \label{dissipator}%
\end{equation}
with $L\overset{\text{def}}{=}\sqrt{\Gamma}\sigma_{-}$ being the Lindblad jump
operator. In Eq. (\ref{dissipator}), $\sigma_{-}\overset{\text{def}}%
{=}\left\vert 0\right\rangle \left\langle 1\right\vert $ and $\sigma
_{+}\overset{\text{def}}{=}\left\vert 1\right\rangle \left\langle 0\right\vert
$ are the lowering and raising operators, respectively. Moreover, $\Gamma$ in
Eq. (\ref{dissipator}) is known as the decay rate, related to the damping
probability $\gamma$ via the relation $\gamma=1-e^{-\Gamma t}$. For a qubit,
the Hamiltonian $\mathrm{H}$ in Eq. (\ref{markov}) is often taken to be the
null operator or $(\omega/2)\sigma_{z}$.

\subsection{Damping, no Hamiltonian dynamics, and time-independent observable}

In the first example, we assume $\mathrm{H}$ in Eq. (\ref{markov}) to be absent. Assuming that the initial state of the qubit is given by%
\begin{equation}
\rho\left(  0\right)  \overset{\text{def}}{=}\left(
\begin{array}
[c]{cc}%
\rho_{00}\left(  0\right)  & \rho_{01}\left(  0\right) \\
\rho_{10}\left(  0\right)  & \rho_{11}\left(  0\right)
\end{array}
\right)  \text{,} \label{rozero}%
\end{equation}
integration of Eq. (\ref{markov}) yields the time-evolved state%
\begin{equation}
\rho(t)=\left(
\begin{array}
[c]{cc}%
\rho_{00}\left(  t\right)  & \rho_{01}\left(  t\right) \\
\rho_{10}\left(  t\right)  & \rho_{11}\left(  t\right)
\end{array}
\right)  =\left(
\begin{array}
[c]{cc}%
\rho_{00}\left(  0\right)  +(1-e^{-\Gamma t})\rho_{11}\left(  0\right)  &
e^{-\frac{\Gamma}{2}t}\rho_{01}\left(  0\right) \\
e^{-\frac{\Gamma}{2}t}\rho_{10}\left(  0\right)  & e^{-\Gamma t}\rho
_{11}\left(  0\right)
\end{array}
\right)  \text{,} \label{evolved}%
\end{equation}
where $\dot{\rho}_{00}=\Gamma\rho_{11}$, $\dot{\rho}_{01}=-\left(
\Gamma/2\right)  \rho_{01}$, $\dot{\rho}_{10}=-\left(  \Gamma/2\right)
\rho_{10}$, and $\dot{\rho}_{11}=-\Gamma\rho_{11}$. From Eq. (\ref{evolved}),
we note that the population in $\left\vert 1\right\rangle $ decays
exponentially with rate $\Gamma$, the coherence decays with rate $\Gamma/2$
and, finally, the lost population moves into the ground state $\left\vert
0\right\rangle $.

Having introduced our noise model, we are now ready to check the validity of
the inequality
\begin{equation}
\left(  \frac{d\sigma_{A}}{dt}\right)  ^{2}\leq2\left[  \left\langle \left(
\frac{\partial A}{\partial t}\right)  ^{2}\right\rangle +\frac{\left[
tr\left(  \frac{d\rho}{dt}(\Delta A)^{2}\right)  \right]  ^{2}%
}{4\sigma_{A}^{2}}\right]  \text{,} \label{ineq}%
\end{equation}
where we assume that $\rho(t)$ is given in Eq. (\ref{evolved}) with
$\rho\left(  0\right)  \overset{\text{def}}{=}\left\vert 1\right\rangle
\left\langle 1\right\vert $ being the excited state and, in addition, we take
the observable $A$ as a non explicitly time-dependent observable given by
$A\overset{\text{def}}{=}\sigma_{z}$. Let us begin by evaluating the LHS of
the inequality in Eq. (\ref{ineq}). We have,%
\begin{equation}
\mathrm{LHS}\overset{\text{def}}{=}\left(  \frac{d\sigma_{A}}{dt}\right)
^{2}=\left(  \frac{1}{2\sigma_{A}}\frac{d\sigma_{A}^{2}}{dt}\right)
^{2}=\frac{1}{4\sigma_{A}^{2}}\left(  \frac{d\sigma_{A}^{2}}{dt}\right)
^{2}\text{,} \label{a1}%
\end{equation}
where the variance $\sigma_{A}^{2}$ is given by%
\begin{equation}
\sigma_{A}^{2}\overset{\text{def}}{=}\left\langle A^{2}\right\rangle
-\left\langle A\right\rangle ^{2}=tr\left[  \rho\left(  t\right)
A^{2}\right]  -tr^{2}\left[  \rho\left(  t\right)  A\right]
=1-\left(  1-2e^{-\Gamma t}\right)  ^{2}\text{.} \label{a2}%
\end{equation}
Substituting Eq. (\ref{a2}) into Eq. (\ref{a1}), we arrive at%
\begin{equation}
\mathrm{LHS}=\Gamma^{2}e^{-\Gamma t}\frac{\left(  1-2e^{-\Gamma t}\right)
^{2}}{1-e^{-\Gamma t}}\text{.} \label{a3}%
\end{equation}
Focusing now on the RHS of the inequality in Eq. (\ref{ineq}) and noting that
$\partial A/\partial t$ vanishes, we have%
\begin{equation}
\mathrm{RHS}=\frac{\left[  tr\left(  \frac{d\rho}{dt}(\Delta
A)^{2}\right)  \right]  ^{2}}{2\sigma_{A}^{2}}\text{,} \label{a4}%
\end{equation}
where $\Delta A\overset{\text{def}}{=}A-\left\langle A\right\rangle
=\sigma_{z}-tr\left[  \rho\left(  t\right)  \sigma_{z}\right]
=\sigma_{z}-\left(  1-2e^{-\Gamma t}\right)  \mathbf{1}$, with $\mathbf{1}$
being the identity operator. Then, after calculating $\Delta A^{2}=\left[
1+\left(  1-2e^{-\Gamma t}\right)  ^{2}\right]  \mathbf{1}-2\left(
1-2e^{-\Gamma t}\right)  \sigma_{z}$, using Eq. (\ref{a2}) along with Eq.
(\ref{evolved}) for evaluating $d\rho/dt$ with $\rho\left(  0\right)
\overset{\text{def}}{=}\left\vert 1\right\rangle \left\langle 1\right\vert $,
the expression for $\mathrm{RHS}$ in Eq. (\ref{a4}) reduces to%
\begin{equation}
\mathrm{RHS}=2\Gamma^{2}e^{-\Gamma t}\frac{\left(  1-2e^{-\Gamma t}\right)
^{2}}{1-e^{-\Gamma t}}\text{.} \label{a5}%
\end{equation}
Finally, comparing Eqs. (\ref{a5}) and (\ref{a3}), we have that $\mathrm{LHS}%
\leq$ $\mathrm{RHS}$. Therefore, our fluctuation growth inequality in Eq.
(\ref{ineq}) is appropriately validated for the specific physical scenario
examined here.

Giving a closer look at the term $\left[  tr\left(  \frac{d\rho}%
{dt}(\Delta A)^{2}\right)  \right]  ^{2}$ in Eq. (\ref{ineq}), we note that
the fluctuation growth inequality can also be checked in an alternative
manner. Indeed, we begin by noting that
\begin{equation}
\frac{d}{dt}tr\left[  \rho\left(  t\right)  (\Delta A)^{2}\right]
=tr\left(  \frac{d\rho\left(  t\right)  }{dt}(\Delta A)^{2}\right)
+tr\left[  \rho\left(  t\right)  \frac{d}{dt}(\Delta A)^{2}\right]
\text{,} \label{here}%
\end{equation}
that is,%
\begin{equation}
tr\left(  \frac{d\rho\left(  t\right)  }{dt}(\Delta A)^{2}\right)
=\frac{d}{dt}tr\left[  \rho\left(  t\right)  (\Delta A)^{2}\right]
-tr\left[  \rho\left(  t\right)  \frac{d}{dt}(\Delta A)^{2}\right]
\text{.}%
\end{equation}
Each one of the terms in Eq. (\ref{here}) has a clear physical interpretation.
Indeed, the term on the left-hand-side is the total derivative of the
uncertainty of the observable $A$. It reflects how the expected uncertainty in
$A$ is changing in time, considering both the system's state and the
observable. The first term on the right-hand-side of Eq. (\ref{here}) measures
how the uncertainty in the observable changes due to the change in the quantum
state itself. Finally, the second term on the right-hand-side of Eq.
(\ref{here}) reflects how the expected uncertainty changes due to the
evolution of the observable itself. In our specific amplitude damping example,
it can be shown that this last term vanishes. Therefore, $\left[
tr\left(  \frac{d\rho}{dt}(\Delta A)^{2}\right)  \right]  ^{2}$
equals $(d\sigma_{A}^{2}/dt)^{2}$ where $\sigma_{A}^{2}\overset{\text{def}}%
{=}\left\langle (\Delta A)^{2}\right\rangle =tr\left[  \rho\left(
t\right)  (\Delta A)^{2}\right]  $. Therefore, using Eqs. (\ref{a1}) and
(\ref{a4}), the fluctuation growth inequality in Eq. (\ref{ineq}) becomes%
\begin{equation}
\frac{(d\sigma_{A}^{2}/dt)^{2}}{4\sigma_{A}^{2}}\leq\frac{(d\sigma_{A}%
^{2}/dt)^{2}}{2\sigma_{A}^{2}}\text{.} \label{tomo}%
\end{equation}
Eq. (\ref{tomo}) is obviously true and leads to the same result as the one
obtained in the inequality $\mathrm{LHS}\leq$ $\mathrm{RHS}$ discussed above
with $\mathrm{RHS}=2\times\mathrm{LHS}$.

\subsection{Damping, Hamiltonian dynamics, and time-dependent observable}

In the second example, we assume $\mathrm{H}$ in Eq. (\ref{markov}) to be
equal to $\mathrm{H}\overset{\text{def}}{=}\left(  \omega/2\right)  \sigma
_{z}$. Assuming that the initial state of the qubit is given as in Eq.
(\ref{rozero}), integration of Eq. (\ref{markov}) yields the time-evolved
state%
\begin{equation}
\rho(t)=\left(
\begin{array}
[c]{cc}%
\rho_{00}\left(  t\right)  & \rho_{01}\left(  t\right) \\
\rho_{10}\left(  t\right)  & \rho_{11}\left(  t\right)
\end{array}
\right)  =\left(
\begin{array}
[c]{cc}%
\rho_{00}\left(  0\right)  +(1-e^{-\Gamma t})\rho_{11}\left(  0\right)  &
e^{-\frac{\Gamma}{2}t}e^{-i\omega t}\rho_{01}\left(  0\right) \\
e^{-\frac{\Gamma}{2}t}e^{i\omega t}\rho_{10}\left(  0\right)  & e^{-\Gamma
t}\rho_{11}\left(  0\right)
\end{array}
\right)  \text{,} \label{evolved2}%
\end{equation}
where $\dot{\rho}_{00}=\Gamma\rho_{11}$, $\dot{\rho}_{01}=-\left(
\Gamma/2+i\omega\right)  \rho_{01}$, $\dot{\rho}_{10}=-\left(  \Gamma
/2-i\omega\right)  \rho_{10}$, and $\dot{\rho}_{11}=-\Gamma\rho_{11}$. We note
that since $\rho\left(  0\right)  \overset{\text{def}}{=}\left\vert
1\right\rangle \left\langle 1\right\vert $, $\rho(t)$ in Eq. (\ref{evolved2})
coincides with $\rho\left(  t\right)  $ in Eq. (\ref{evolved}). For other
initial conditions (i.e., $\rho\left(  0\right)  \neq\left\vert 1\right\rangle
\left\langle 1\right\vert $), however, they generally describe distinct
temporal behaviors of the qubit density operator. Having introduced our noise
model, we are now ready to check the validity of the fluctuation growth
inequality in Eq. (\ref{ineq}). In this second example, we assume $A$ to be an
explicitly time-dependent observable given by $A\left(  t\right)
\overset{\text{def}}{=}\cos(t)\sigma_{x}+\sin(t)\sigma_{y}$. Setting
$\xi\left(  t\right)  \overset{\text{def}}{=}\left\langle A\right\rangle
=tr\left[  \rho\left(  t\right)  A\left(  t\right)  \right]  $, we
have $\sigma_{A}^{2}=1-\xi^{2}\left(  t\right)  $. Therefore, the
left-hand-side of the inequality in Eq. (\ref{ineq}) becomes%
\begin{equation}
\mathrm{LHS}=\frac{\xi^{2}(t)\dot{\xi}^{2}(t)}{1-\xi^{2}(t)}\text{,}
\label{say1}%
\end{equation}
where $\dot{\xi}\overset{\text{def}}{=}d\xi/dt$. Moreover, we note that
$\left\langle \left(  \partial A/\partial t\right)  ^{2}\right\rangle =1$ and,
in addition, $(\Delta A)^{2}=\left[  1+\xi^{2}(t)\right]  \mathbf{1-}%
2\xi\left(  t\right)  A$. Therefore, we have%
\begin{equation}
\left[  tr\left(  \frac{d\rho}{dt}(\Delta A)^{2}\right)  \right]
^{2}=\left[  1+\xi^{2}(t)\right]  tr\left(  \frac{d\rho}{dt}\right)
-2\xi(t)tr\left(  \frac{d\rho}{dt}A\right)  =-2\xi(t)\dot{\xi}\left(
t\right)  \text{,} \label{phil}%
\end{equation}
since $tr\left(  d\rho/dt\right)  =d\left[  tr\left(
\rho\right)  \right]  /dt=0$ and $tr\left[  (d\rho/dt)A\right]
=d\left[  tr\left(  \rho A\right)  \right]  /dt-tr\left[
\rho(dA/dt)\right]  =d\left\langle A\right\rangle /dt$ because $\left\langle
dA/dt\right\rangle =tr\left[  \rho(dA/dt)\right]  =0$ in our case.
Therefore, recalling that $\sigma_{A}^{2}=1-\xi^{2}\left(  t\right)  $ and
using Eq. (\ref{phil}), the right-hand-side of the inequality in Eq.
(\ref{ineq}) reduces to%
\begin{equation}
\mathrm{RHS}=2\left[  1+\frac{\xi^{2}(t)\dot{\xi}^{2}(t)}{1-\xi^{2}%
(t)}\right]  \text{.} \label{say2}%
\end{equation}
Finally, comparing Eqs. (\ref{say2}) and (\ref{say1}), we have that
$\mathrm{LHS}\leq$ $\mathrm{RHS}$. As a side remark, we observe that
$\xi\left(  t\right)  =0$ in our case, so the inequality reduces to $0\leq2$.
This confirms that our fluctuation growth inequality in Eq. (\ref{ineq}) holds
true for the second physical scenario as well.

\section{Conclusion}
We considered an open quantum system in two cases to find an upper bound for the rate of fluctuation growth of an observable $A$.  We first looked at an open system with an evolution operator that was obtained via two ways; using the Taylor expansion for the generator of evolutions, given by Eq. (\ref{eq:11}), and using the time-ordered expansion under a time-dependent Hamiltonian.  The evolution operator was the same, as expected, in both methods.  Given the specific form of the pseudo-Hamiltonian associated with the evolution operator, we found the rate of change of the average value of observable $A$, given by Eq. (\ref{eq:17}), as well as the time derivative of the average value of $A^{2}$, shown in Eq. (\ref{eq:23}).  Using these results we found the time derivative of the uncertainty of the observable, $\sigma_{A}$.  Then using the Cauchy-Schwarz inequality along with another general inequality that is always true for real numbers given by Eq. (\ref{eq:41}), we found that the rate of change of the variance of observable $A$ is upper bounded by the average value of its time derivative, plus rate at which uncertainty in observable $A$ changes due to the dynamics of the open system,  as expressed by Eq. (\ref{eq:46}).   

We then considered a general case for an open quantum system where no information about the evolution is known.  The time derivative of the variance of $A$ was found, given by Eq. (\ref{eq:50}).  Then using the Cauchy-Schwarz inequality along with the same general inequality for real numbers used in the previous case with known evolution operator, we found that the rate of change of fluctuations of observable $A$ is upper bounded by the average value of the time derivative of the observable $A$, plus the rate of change of uncertainty of the observable due to the dynamics of the open system, given by Eq. (\ref{eq:62}).  Both inequalities simply indicate that the rate of change of uncertainty of an observable in an open quantum system over time is bounded by how fast the operator as well as the state change.  We then considered two examples and verified the inequality in both cases where the Hamiltonian was known and when no information about the Hamiltonian was considered with time-dependent and time-independent observables.

Comparing the results from Eqs. (\ref{eq:46}) and (\ref{eq:62}) with the inequality derived in \cite{H}, which is given by 
\begin{equation}
\label{eq:63}
\dot{\sigma}^{2}_{A} \leq \sigma^{2}_{\dot{A}},
\end{equation} 
suggests that Eqs. (\ref{eq:46}), (\ref{eq:62}) seems to be looser bounds compared to the inequality given by Eq. (\ref{eq:63}), derived in Refs. \cite{H}, \cite{CRBC}.
A similar inequality is derived on quantum acceleration limit \cite{P, AC, CCBA}. 
We show how the bound found for closed systems in Ref. \cite{CRBC} is recovered when the symmetric definition of covariance is used to derive the inequality in Appendix A, and when the nonsymmetric definition of the covariance is used in Appendix B.

The fact that Eqs. (\ref{eq:46}), (\ref{eq:62}) seem to be looser bounds compared to Eq. (\ref{eq:63}) is expressed with an explicit example, see Appendix C.  Moreover, the bounds naturally arise when one includes evolution of the density operator and/or the generator of evolution $\Omega$.  This suggests that when we separate the contributions of the dynamics of the observable $\partial_{t}A$ versus the change of state $\dot{\rho}$ the bound seems to become looser.  In other words, trying to use more detailed contributions through splitting the terms and bounding each separately seems to result in mathematically weaker bounds, while the bound given by Eq. (\ref{eq:63}) keeps all correlations between terms with all fluctuations encoded in $\sigma_{\dot{A}}$.     

The fact that we obtain the same inequality whether or not the pseudo-Hamiltonian is known, indicates that the bound is independent of the generator.  That is, the result found here does not depend on a specific form of a generator.  Therefore, the maximum rate at which uncertainty of an observable can evolve is set regardless of any specific generator.  Indeed experimentally it is specifically important and more accessible that one does not need to know the full evolution law for open systems.  This can be a useful tool for designing controls that stay within dynamic variance limits and controlling noise, which are crucial in realistic systems that are modeled as open systems in quantum mechanics.

The expectation of looser bound  for an open system compared to a closed system can be justified since in closed systems, fluctuations are due to intrinsic uncertainties and noncommutativity of operators.  However, in open systems, there is additional noise induced by the environment that contributes to the variance.  Examples of measuring such fluctuations can be found in different areas of quantum studies. To give a few examples, in Refs. \cite{LBMW, GMEHB} position and momentum fluctuations have been measured, while in Ref. \cite{CRBM} fluctuations of spin have been measured, and in Ref. \cite{P} quantum thermodynamic fluctuations are measured.

\section*{Acknowledgements}
Any opinions, findings and conclusions or recommendations expressed in this material are those of the author(s) and do not necessarily reflect the views of their home institutions.

\bigskip

\bigskip

\bigskip

\pagebreak

\section*{Appendix A. Eq. (69) for Closed Systems}

In this Appendix, we show that Eq. (\ref{eq:62}) which is the bound found when no information about the evolution is known using the nonsymmetric definition of covariance, reduces to Eq. (\ref{eq:63}) when the system is closed.  In Ref. \cite{H} the inequality 
\begin{equation}
\label{eq:H1}
\left| \dot{\sigma}_{A}\right| \leq \sigma_{\dot{A}}
\end{equation}
is proven by showing that the following equation is true 
\begin{equation}
\label{eq:H2}
\frac{d\left\langle \Delta A^{2}\right\rangle }{dt}= 2\left\langle \Delta A , \Delta \dot{A}\right\rangle ,
\end{equation}
where $\Delta \dot{A} \stackrel{\text{def}}{=} \dot{A}-\left\langle \dot{A}\right\rangle $.  Then using the Cauchy-Schwarz inequality, one can write
\begin{equation}
\label{eq:H3}
\left| \left\langle \Delta A , \Delta \dot{A}\right\rangle \right| \leq \sqrt{\left\langle \Delta A^{2}\right\rangle }\sqrt{\left\langle \Delta \dot{A}^{2}\right\rangle }=\sigma_{A} \sigma_{\dot{A}},
\end{equation} 
that is
\begin{equation}
\label{eq:H4}
\left| \left\langle \Delta A , \Delta \dot{A}\right\rangle \right| \leq \sigma_{A} \sigma_{\dot{A}}.
\end{equation} 
Now using 
\begin{equation}
\label{eq:H5}
\frac{d\left\langle \Delta A^{2}\right\rangle }{dt}= \frac{d \sigma^{2}_{A}}{dt} = 2 \sigma_{A} \dot{\sigma}_{A}
\end{equation}
along with Eq. (\ref{eq:H4}), one obtains 
\begin{equation}
\label{eq:H6}
\left| \sigma_{A} \dot{\sigma}_{A} \right| \leq \sigma_{A} \sigma_{\dot{A}},
\end{equation} 
which gives the inequality given by Eq. (\ref{eq:H1}), or Eq. (\ref{eq:63}).

In order to show that Eq. (\ref{eq:H2}) is true, it is shown in Ref. \cite{H} that both sides of this equation equal the following term
\begin{equation}
\label{eq:H7}
\left( \left\lbrace \Delta A, \partial_{t}A \right\rbrace + \mathcal{L}^{\dagger}[A^{2}]-2\left\langle A\right\rangle \mathcal{L}^{\dagger}[A]|\rho\right), 
\end{equation}
where $\mathcal{L}^{\dagger}[.]$ is the dual map of the Lindblad operator $\mathcal{L}$ such that $(A|\mathcal{L}[\rho])=(\mathcal{L}^{\dagger}[A]|\rho)$, and $\dot{\rho}=\mathcal{L}[\rho]$.
A detailed proof is also given in \cite{CRBC}.

Next, considering the result given by Eq. (\ref{eq:62}) with no information about a pseudo-Hamiltonian, we found for open systems 
\begin{equation}
\label{eq:H8}
\frac{d\sigma^{2}_{A}}{dt}=tr \left( \dot{\rho}\Delta A^{2}\right)  + 2Re Cov(A,\partial_{t}A)= tr \left( \dot{\rho}\Delta A^{2}\right)  + \left\langle \left\lbrace \Delta A, \partial_{t}A\right\rbrace \right\rangle .
\end{equation}
Rewriting the first term on the RHS of Eq. (\ref{eq:H8}) gives  
\begin{equation}
\label{eq:H9}
tr(\dot{\rho}\Delta A^{2})=tr(\mathcal{L}[\rho]\Delta A^{2}) =tr(\rho \mathcal{L}^{\dagger}[\Delta A^{2}]).
\end{equation}
Knowing that $\Delta A^{2}=(A-\left\langle A\right\rangle)^{2}=A^{2}-2\left\langle A\right\rangle A+\left\langle A\right\rangle ^{2}$, we can write
\begin{equation}
\label{eq:H10}
\mathcal{L}^{\dagger}[\Delta A^{2}]=\mathcal{L}^{\dagger}[A^{2}]-2\left\langle A\right\rangle \mathcal{L}^{\dagger}[A]=\mathcal{L}^{\dagger}[A^{2}]-2\left\langle A\right\rangle \mathcal{L}^{\dagger}[A]
\end{equation}
with $\mathcal{L}^{\dagger}[I]=0$, where $I$ is the identity matrix.  Then Eq. (\ref{eq:H8}) can be rewritten as
\begin{equation}
\label{eq:H11}
\frac{d\sigma^{2}_{A}}{dt}=\left\langle \mathcal{L}^{\dagger}[A^{2}]-2\left\langle A\right\rangle \mathcal{L}^{\dagger}[A] + \left\lbrace \Delta A, \partial_{t}A\right\rbrace \right\rangle 
\end{equation}
which is the same term found in Eq. (\ref{eq:H7}) in Ref. \cite{H}.

Next, we need to show that the RHS of Eq. (\ref{eq:H2}) is also equal to Eq. (\ref{eq:H7}), that is
\begin{equation}
\label{eq:H12}
2\left\langle \Delta A,\Delta \dot{A}\right\rangle = \left( \left\lbrace \Delta A, \partial_{t}A \right\rbrace + \mathcal{L}^{\dagger}[A^{2}]-2\left\langle A\right\rangle \mathcal{L}^{\dagger}[A]|\rho\right).
\end{equation}
Knowing that 
\begin{equation}
\label{eq:H12.1}
\left\langle \Delta A,\Delta \dot{A}\right\rangle =\left\langle \left( A-\left\langle A\right\rangle \right), \left( \dot{A}-\left\langle \dot{A} \right\rangle\right)\right\rangle  = \left\langle A\dot{A}\right\rangle  - \left\langle A\right\rangle \left\langle \dot{A}\right\rangle
\end{equation}
with $\left\langle \Delta A\right\rangle =\left\langle \Delta \dot{A}\right\rangle =0$, one can see that 
\begin{equation}
\label{eq:H12.2}
\left\langle \Delta A, \dot{A}\right\rangle =\left\langle \left( A-\left\langle A\right\rangle \right),\dot{A}\right\rangle =\left\langle A\dot{A}\right\rangle- \left\langle A\right\rangle \left\langle \dot{A}\right\rangle
\end{equation}
that is, $\left\langle \Delta A,\Delta\dot{A}\right\rangle =\left\langle \Delta A,\dot{A}\right\rangle $. 
From Eqs. (\ref{eq:H12.1}) and (\ref{eq:H12.2}), the relation we need to show is true will then be 
\begin{equation}
\label{eq:H12.5}
\left\langle \left\lbrace \Delta A,\dot{A}\right\rbrace \right\rangle = \left( \left\lbrace \Delta A, \partial_{t}A \right\rbrace + \mathcal{L}^{\dagger}[A^{2}]-2\left\langle A\right\rangle \mathcal{L}^{\dagger}[A]|\rho\right).
\end{equation}  
Considering a closed system, we can write $\dot{A}=\partial_{t}A+i[H,A]$, which allows us to write
\begin{equation}
\begin{aligned}
\left\langle \left\lbrace \Delta A,\dot{A}\right\rbrace \right\rangle &= \left\langle \left\lbrace \Delta A,\left( \partial_{t}A+i[\mathrm{H},A]\right) \right\rbrace \right\rangle \\
&= \left\langle \left\lbrace \Delta A,\partial_{t}A\right\rbrace \right\rangle +\left\langle \left\lbrace \Delta A,i[\mathrm{H},A]\right\rbrace \right\rangle \\
&= \left\langle \left\lbrace \Delta A,\partial_{t}A\right\rbrace \right\rangle +\left\langle \left\lbrace (A-\left\langle A\right\rangle ),i[\mathrm{H},A]\right\rbrace \right\rangle \\
&= \left\langle \left\lbrace \Delta A,\partial_{t}A\right\rbrace \right\rangle +\left\langle \left\lbrace A,i[\mathrm{H},A]\right\rbrace \right\rangle -2\left\langle A \right\rangle \left\langle i[\mathrm{H},A]\right\rangle . 
\end{aligned}
\label{eq:H13}
\end{equation}
Using $\mathcal{L}^{\dagger}[A^{2}]=\left\lbrace A,\mathcal{L}^{\dagger}[A]\right\rbrace $ and $\mathcal{L}^{\dagger}[A]=i[H,A]$, we can rewrite the last line of Eq. (\ref{eq:H13}) as
\begin{equation}
\label{eq:H14}
\left\langle \left\lbrace \Delta A, \dot{A}\right\rbrace \right\rangle =\left\langle \left\lbrace \Delta A,\partial_{t}A\right\rbrace +\mathcal{L}^{\dagger}[A^{2}]-2\left\langle A\right\rangle \mathcal{L}^{\dagger}[A]\right\rangle ,
\end{equation}
which is the same result found in Eq. (\ref{eq:H11}).  Equating Eqs. (\ref{eq:H11}) and (\ref{eq:H14}) yields
\begin{equation}
\label{eq:H15}
\frac{d\sigma^{2}_{A}}{dt}=\left\langle \left\lbrace \Delta A,\dot{A}\right\rbrace \right\rangle 
\end{equation}
 in a closed system.
  
This indicates that the result valid for open systems found in Eq. (\ref{eq:62}) does give the limit found in Ref. \cite{CRBC} for closed systems.

\section*{Appendix B. Eq. (77) for Closed Systems} 

In this Appendix we will show how the inequality found for open systems with no information about the pseudo-Hamiltonian using the symmetric definition for covariance, reduces to the inequality found in Ref. \cite{CRBC} for closed systems given by Eq. (\ref{eq:63}).  Considering Eq. (\ref{eq:P4.8}) along with $\sigma^{2}_{\partial_{t}A}=\left\langle (\partial_{t}A)^{2}\right\rangle -\left\langle \partial_{t}A\right\rangle ^{2}$, yields
\begin{equation}
\label{eq:A1}
\dot{\sigma}_{A} \leq \frac{tr(\dot{\rho}\Delta A^{2})}{2\sigma_{A}}+\sqrt{\left\langle (\partial_{t}A)^{2}\right\rangle } \equiv \frac{tr(\dot{\rho}\Delta A^{2})}{2\sigma_{A}} + \sqrt{\sigma^{2}_{\partial_{t}A}+\left\langle \partial_{t}A\right\rangle ^{2}}. 
\end{equation}
Considering Eq. (\ref{eq:A1}), if $\left\langle \partial_{t}A\right\rangle ^{2} =0$, and $\Delta A^{2}\varpropto I$, (e.g $\Delta A^{2}=f(t) I$), then Eq. (\ref{eq:A1}) becomes
\begin{equation}
\label{eq:A2}
\frac{f(t)}{2\sigma_{A}}tr(\dot{\rho})+\sqrt{\sigma^{2}_{\partial_{t}A}+\left\langle \partial_{t}A\right\rangle ^{2}}=\sqrt{\sigma^{2}_{\partial_{t}A}}.
\end{equation}
Squaring both sides of Eq. (\ref{eq:A2}), and setting $f(t)=1$ gives the inequality found for closed systems given by Eq. (\ref{eq:63}), where $\sigma_{\dot{A}}=\sigma_{\partial_{t}A}$.

\section*{Appendix C.  On the Inequality $\dot{\sigma}_{A}^{2}\leq\sigma_{\dot{A}}^{2}$}

In this Appendix, we wish to verify if $\dot{\sigma}_{A}^{2}\leq\sigma
_{\dot{A}}^{2}$ holds in our first illustrative example where we consider a
qubit undergoing pure amplitude damping noise, with initial state given by
$\rho\left(  0\right)  \overset{\text{def}}{=}\left\vert 1\right\rangle
\left\langle 1\right\vert $ and, lastly, with an observable $A\overset
{\text{def}}{=}\sigma_{z}$. Under this nonunitary dynamical evolution, we
recall from the main part of the manuscript that the evolved state
$\rho\left(  t\right)  $ is given by%
\begin{equation}
\rho\left(  t\right)  =\left(
\begin{array}
[c]{cc}%
\gamma\left(  t\right)   & 0\\
0 & 1-\gamma\left(  t\right)
\end{array}
\right)  =\left(
\begin{array}
[c]{cc}%
1-e^{-\Gamma t} & 0\\
0 & e^{-\Gamma t}%
\end{array}
\right)  \text{.}\label{news1}%
\end{equation}
Using Eq. (\ref{news1}), we observe that%
\begin{align}
\sigma_{A}^{2} &  =\left\langle A^{2}\right\rangle -\left\langle
A\right\rangle ^{2}\nonumber\\
&  =tr\left[  \rho\left(  t\right)  A^{2}\right]  -tr^{2}\left[  \rho\left(
t\right)  A\right]  \nonumber\\
&  =1-\left[  2\gamma\left(  t\right)  -1\right]  ^{2}\nonumber\\
&  =4\gamma\left(  t\right)  \left[  1-\gamma\left(  t\right)  \right]
\text{,}%
\end{align}
that is, after some algebra,%
\begin{equation}
\left\vert \frac{d\sigma_{A}}{dt}\right\vert =\frac{\left\vert 1-2\gamma
\left(  t\right)  \right\vert \cdot\left\vert \dot{\gamma}(t)\right\vert
}{\sqrt{\gamma\left(  t\right)  \left[  1-\gamma\left(  t\right)  \right]  }%
}\text{.}\label{news2}%
\end{equation}
To calculate $\sigma_{\dot{A}}^{2}$, we begin by using the Lindblad adjoint to
express $\dot{A}$ as $\partial_{t}A+i\left[  \mathrm{H}\text{, }A\right]
+\mathcal{D}^{\dagger}\left(  A\right)  $ \cite{gardiner, petruccione}. Since $A$ does not depend
explicitly on time and since the Hamiltonian dynamics is absent in our case,
we have $\partial_{t}A=0$ and $i\left[  \mathrm{H}\text{, }A\right]  =0$.
Therefore, $\dot{A}$ reduces to%
\begin{equation}
\dot{A}=\frac{dA}{dt}=\mathcal{D}^{\dagger}\left(  A\right)  =L^{\dagger
}AL-\frac{1}{2}\left\{  L^{\dagger}L\text{, }A\right\}  \text{,}\label{news3}%
\end{equation}
with $L\overset{\text{def}}{=}\sqrt{\Gamma}\sigma_{-}=\sqrt{\Gamma}\left\vert
0\right\rangle \left\langle 1\right\vert $ and $L^{\dagger}\overset
{\text{def}}{=}\sqrt{\Gamma}\sigma_{+}=\sqrt{\Gamma}\left\vert 1\right\rangle
\left\langle 0\right\vert $. A simple calculation yields, $\dot{A}%
=2\Gamma\left\vert 1\right\rangle \left\langle 1\right\vert $. Therefore, we
arrive at%
\begin{align}
\sigma_{\dot{A}}^{2} &  =\left\langle \dot{A}^{2}\right\rangle -\left\langle
\dot{A}\right\rangle ^{2}\nonumber\\
&  =tr\left[  \rho\left(  t\right)  \dot{A}^{2}\right]  -tr^{2}\left[
\rho\left(  t\right)  \dot{A}\right]  \nonumber\\
&  =4\Gamma^{2}\left[  1-\gamma\left(  t\right)  \right]  -\left\{
2\Gamma\left[  1-\gamma\left(  t\right)  \right]  \right\}  ^{2}\nonumber\\
&  =4\Gamma^{2}\gamma\left(  t\right)  \left[  1-\gamma\left(  t\right)
\right]  \text{,}%
\end{align}
that is,%
\begin{equation}
\sigma_{\dot{A}}=2\Gamma\sqrt{\gamma\left(  t\right)  \left[  1-\gamma\left(
t\right)  \right]  }\text{.}\label{news4}%
\end{equation}
Combining Eqs. (\ref{news2}) and (\ref{news4}), we have that $\left\vert
\dot{\sigma}_{A}\right\vert \leq\sigma_{\dot{A}}$ if and only if%
\begin{equation}
\frac{\left\vert 1-2\gamma\left(  t\right)  \right\vert \cdot\left\vert
\dot{\gamma}(t)\right\vert }{\sqrt{\gamma\left(  t\right)  \left[
1-\gamma\left(  t\right)  \right]  }}\leq2\Gamma\sqrt{\gamma\left(  t\right)
\left[  1-\gamma\left(  t\right)  \right]  }\text{.}\label{news5}%
\end{equation}
Since $\gamma\left(  t\right)  =1-e^{-\Gamma t}$, $\dot{\gamma}(t)=\Gamma
e^{-\Gamma t}=\Gamma\left[  1-\gamma\left(  t\right)  \right]  $. Therefore,
after some algebraic manipulations and assuming $\gamma\left(  t\right)
\neq1$, the inequality in Eq. (\ref{news5}) reduces to%
\begin{equation}
\left\vert 1-2\gamma\left(  t\right)  \right\vert \leq2\gamma\left(  t\right)
\text{.}\label{news6}%
\end{equation}
The inequality in Eq. (\ref{news6}) is satisfied only if $\gamma\left(
t\right)  \geq1/4$ with $\gamma\in\left(  0\text{, }1\right]  $. We therefore
conclude that the inequality $\dot{\sigma}_{A}^{2}\leq\sigma_{\dot{A}}^{2}$
holds true only when $\gamma\left(  t\right)  \geq1/4$, given this specific
noise model, this particular initial condition, and this selected observable.
A graphical visualization of the inequality $\dot{\sigma}^{2}_{A} \leq \sigma^{2}_{\dot{A}}$ appears in Fig. 1.

\begin{figure}[t]
\centering
\includegraphics[width=0.5\textwidth] {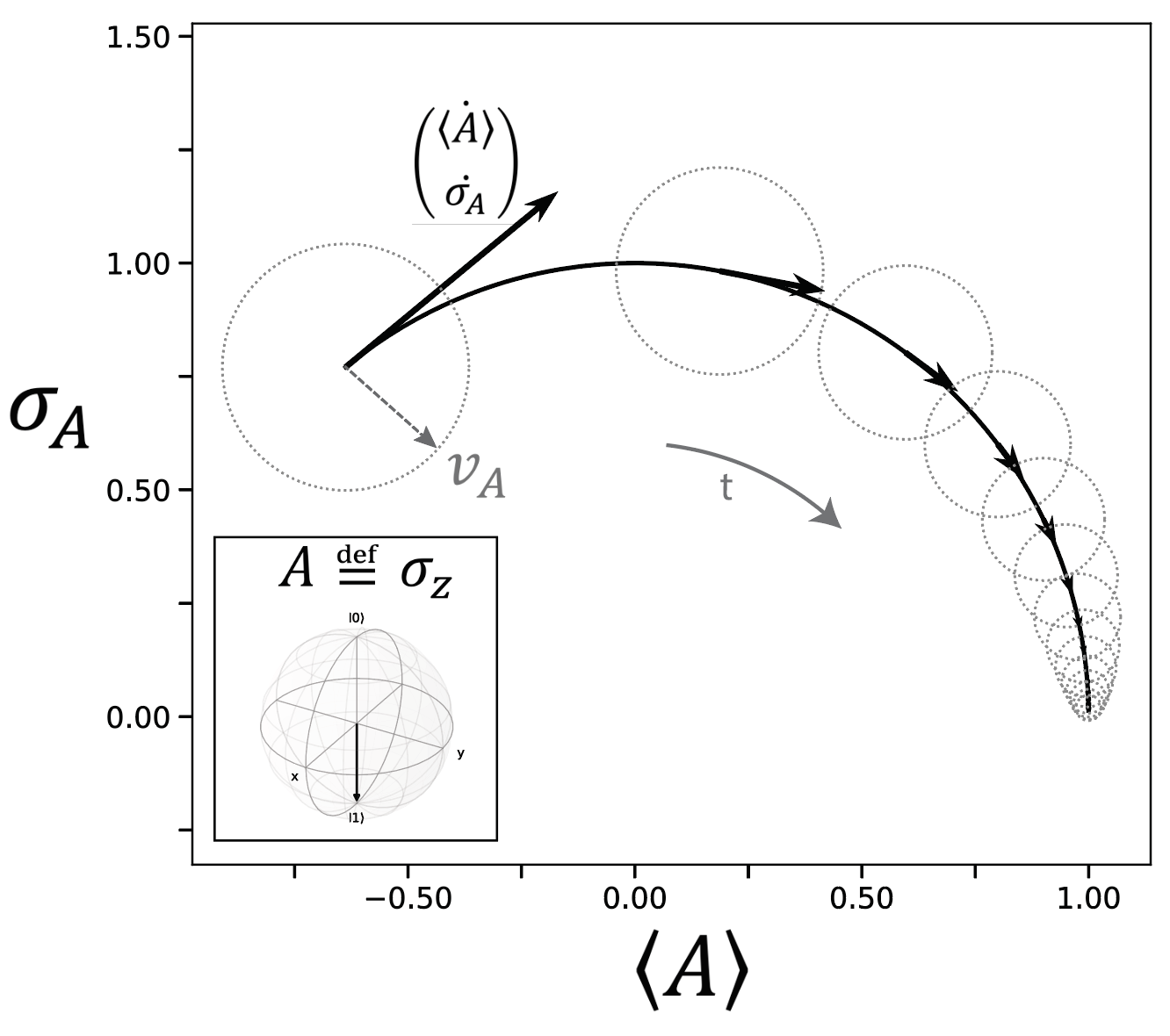}\caption{A graphical representation illustrating the temporal progression of the
inequality $\dot{\sigma}_{A}^{2}\leq\sigma_{\dot{A}}^{2}$ (or, alternatively,
$\dot{\mu}_{A}^{2}+\dot{\sigma}_{A}^{2}\leq v_{A}^{2}$ with $\dot{\mu}_{A}%
^{2}\overset{\text{def}}{=}\left\langle \dot{A}\right\rangle ^{2}$ and
$v_{A}^{2}\overset{\text{def}}{=}\left\langle \dot{A}^{2}\right\rangle $) in
the $\left(  \mu_{A}\text{, }\sigma_{A}\right)  $-plane where $\mu_{A}%
\overset{\text{def}}{=}\left\langle A\right\rangle $ and $A\overset
{\text{def}}{=}\sigma_{z}$. In our example, we have $\mu_{A}\left(  t\right)
=1-2e^{-\Gamma t}$, $\sigma_{A}\left(  t\right)  =2e^{-\frac{\Gamma}{2}t}%
\sqrt{1-e^{-\Gamma t}}$, and $v_{A}\left(  t\right)  =2\Gamma e^{-\frac
{\Gamma}{2}t}$. The initial state of the system is assumed to be $\rho\left(
0\right)  =\left\vert 1\right\rangle \left\langle 1\right\vert $. From our
analysis, it happens that the inequality is violated (satisfied) for
$\gamma\left(  t\right)  <1/4$ ($\gamma\left(  t\right)  \geq1/4$ ) or, in
terms of time, for $t<(1/\Gamma)\ln(4/3)$ ($t\geq(1/\Gamma)\ln(4/3)$).
Clearly, $\gamma\left(  t\right)  $ and $\Gamma$ are the damping probability
and the decay rate, respectively. From a graphical standpoint, this violation
is visualized in terms of a bold arrow of length $\left\langle \dot
{A}\right\rangle ^{2}+\dot{\sigma}_{A}^{2}$ which is outside the circle of
radius $v_{A}$. Moreover, when the inequality holds, the bold arrow is
contained within the circle. Finally, observe that when $t$ approaches
infinity, $\mu_{A}$ approaches one, $\sigma_{A}$ converges to zero, and the
density operator of the system transitions to $\left\vert 0\right\rangle
\left\langle 0\right\vert $.}%
\end{figure}

\end{document}